\documentclass[twocolumn,amsmath,amssymb]{revtex4}
\usepackage{epsfig}
\usepackage{graphicx}

\begin{document}

\title{Reply to Sergiu I. Vacaru's ``Critical remarks on Finsler modifications of gravity and cosmology
by Zhe Chang and Xin Li"}

\author{Xin Li$^{1,3}$}
\email{lixin@itp.ac.cn}
\author{Zhe Chang$^{2,3}$}
\email{changz@ihep.ac.cn}
\affiliation{${}^1$Institute of Theoretical Physics,
Chinese Academy of Sciences, 100190 Beijing, China\\
${}^2$Institute of High Energy Physics, Chinese Academy
of Sciences, 100049 Beijing, China\\
${}^3$Theoretical Physics Center for Science Facilities, Chinese Academy of Sciences}

\begin{abstract}
This is our reply to "Critical remarks on Finslerian modifications of
gravity and cosmology by Zhe Chang and Xin Li", Sergiu I. Vacaru,
Phys. Lett. B 690 (2010) 224. It is pointed out that the Finslerian
modifications of gravity and cosmology (Zhe Chang and Xin Li, Phys.
Lett. B 676 (2009) 173; {\it ibid} 668 (2008) 453) is a suggestion
on the generalization of Einstein's gravity and cosmology, but not a
proof for theorems in geometry. False or true of the theory should
be tested by experiments or observations. We show that the arguments of
Sergiu I. Vacaru were based a wrong logic. A personal claim can not
be used to prove any other theory be wrong. To get the claim: {\it
``we may construct more ``standard" physical Finsler
classical/quantum gravity theories for metric compatible connections
like the Cartan d-connection"} , Sergiu I. Vacaru should complete a
consistent presentation at least. We suggest Sergiu I. Vacaru to make
some predictions on gravity and cosmoligy using his {\it ``standard"
physical Finsler classical/quantum gravity theories} as we did, and
compare them with astronomical observations. By the way, we should
say that it is still really far from a theory of quantum gravity.
\end{abstract}

\maketitle

Recently, Sergiu I. Vacaru published critical remarks\cite{Vacaru}
on our work {\it Finslerian modifications of gravity and cosmology}
\cite{Finsler DE,Finsler DM}. First of all, we thank Vacaru for
paying attention to our researches. We are happy to read any kind of
criticisms and comments on the papers. The Finslerian modifications of gravity and cosmology are really not complete and
in the course of development.  However, we found that the comments
of Sergiu I. Vacaru was based a wrong logic.   The Finslerian
modifications of gravity and cosmology is a suggestion on the
generalization of Einstein's gravity and cosmology, but not a proof
for theorems in geometry. False or true of the theory should be
tested by experiments or observations. A personal claim can not be
used to prove any other researches to be wrong. To get the claim: {\it ``we
may construct more ``standard" physical Finsler classical/quantum
gravity theories for metric compatible connections like the Cartan
d-connection"} , Sergiu I. Vacaru should complete a consistent
presentation at least. We suggest Sergiu I. Vacaru to make some
predictions on gravity and cosmoligy using his {\it ``standard"
physical Finsler classical/quantum gravity theories} as we did, and
compare them with astronomical observations.

In the following, we will reply to some key points in the comments of
Sergiu I. Vacaru.\\

Question1: {\it ``The Chern connection is not metric compatible and
not the unique connection in Finsler geometry".}\\

Reply: It is correct. The Chern connection is not metric compatible.
The statement we gave in \cite{Finsler DM} should be replaced by a
more clear presentation. In fact, in the second paper of the
series\cite{Finsler DE}, just before the formula (2), we have already
altered the statement as: ``In Finsler manifold, there exists a
unique linear connection-the Chern connection. It is torsion
freeness and almost metric compatibility". Here the word ``unique"
just means that the Chern connection is determined by the conditions(or structural equation) of
torsion freeness and almost metric compatibility. It should not be
read as that the Chern connection is the unique connection in
Finsler geometry.\\

Question2: {\it ``The metric incompatibility make more difficult the
definition of spinors and conservation laws in Finsler gravity and
does not allow ``simple" (super) string and noncommutative
generalizations like we proposed."}\\

Reply: Our paper just presented a classical modification of gravity
and cosmology and did not concern any aspect of quantum theory of
gravity. Therefore, we do not think it is a comment on our papers.
It is strange that a claim on quantum gravity can be used to
criticize a classical theory of gravity. We should say that here the
logic of Sergiu I. Vacaru is wrong. Even though, we still would like
to point out that the comment made a strong conclusion without any
proof. It is a pity that a proof of the no go theorem can not be
found in the comments\cite{Vacaru}.\\

Question3: {\it ``The Ricci tensor introduced by H. Akbar-Zadeh
\cite{Akbar} is not correct for all Finsler geometry/gravity models.
There were considered various types of Ricci type tensors in Finsler
geometries."}\\

Reply: We do not know any mathematician has presented the theorem.
Sergiu I. Vacaru did not give any sound proof about his
assertion either. In fact, these various types of Ricci tensors in Finsler
geometry and the so-called ``gravitational field equations" constructed
by them depend on the chosen connection. It implies that
different gravitational field equations could be obtained while one uses
different connections to calculate it, even all the connections are
metric compatible. This brings up the problem that which
gravitational field equation is the physical one. On the contrary,
the Ricci tensor that introduced by  Akbar-Zadeh \cite{Akbar} does
not face such a problem. It is given as
\begin{equation}
Ric_{\mu\nu}=\frac{\partial^2\left(\frac{1}{2}F^2Ric\right)}{\partial y^\mu\partial y^\nu},
\end{equation}
where the Finsler metric is defined as $g_{\mu\nu}\equiv\frac{\partial^2\left(\frac{1}{2}F^2\right)}{\partial y^\mu\partial y^\nu}$, and the Ricci scalar ``$Ric$" is the trace of the predecessor of the flag curvature. The flag curvature \cite{Book by Bao} in Finsler geometry is the counterpart of the sectional curvature in Riemannian geometry. It is a geometrical invariant. Furthermore, the same flag curvature is obtained for any connection that chosen in Finsler space. Thus, the same Ricci tensor is obtained for any connection that chosen in Finsler space Therefore, the Ricci tensor introduced by Akbar-Zadeh \cite{Akbar} is a reasonable and well-defined one.\\

Question4: {\it ``Sergiu I. Vacaru claimed that he may construct more ``standard" physical Finsler
classical/quantum gravity theories for metric compatible connections like the Cartan d-connection".}\\

Reply: To become a theory, a claim should be complete and consistent
at least. The most important thing for a physical theory is that it makes
predictions that can be tested through experiments and observations.
To our point of view, Sergiu I. Vacaru's Finsler gravity theories
are still premature. Newton's theory of gravity can not be used to
kill Einstein's general relativity. Einstein's equations of
gravitational field can not be used to kill the hypothesis of dark
energy and dark matter. 
Another reason for the prematureness is that the Einstein's tensor $E(\hat{D})$ in Sergiu I. Vacaru's Finsler gravity theories is not a conserved quantity \cite{Vacaru1} (in the sense of covariant differentiation). This is
also pointed out by Sergiu I. Vacaru himself in his comments (see
the footnote 3 and the formula (3) in \cite{Vacaru}). It is
well-known that in general relativity the Einstein tensor is a
conserved quantity. The Einstein's gravitational field equation
constructed in such a form due to the requirement that the
energy-momentum tensor must conserve. And this conservation law of
energy-momentum tensor is of vital importance and has been
extensively used and embedded in different branches of modern
physics. Any theory that does not subject to this rule can not be
recognized as a physical one. Neither can Sergiu I. Vacaru's Finsler
gravity theories. At least, Sergiu I. Vacaru should give a conserved
quantity which is the counterpart of energy-momentum tensor in the
framework of Finsler geometry.   Sergiu I. Vacaru has claimed that
the quantum gravity theory is ``almost sure" of generalized Finsler
type. We wish him publish his claim in an isolated paper. To discuss
details on this topic is out of range of our reply.

\bigskip

\centerline{\large\bf Acknowledgements} \vspace{0.5cm}
 We would like to thank Prof. Z. Shen and Dr. M. Li for useful discussions. The
work was supported by the NSF of China under Grant No. 10525522 and
10875129.

\end{document}